%% file: paper.tex
\documentclass[letterpaper,twocolumn,10pt]{article}

\def\tool{\textsc{Hydra}\xspace}
\def\papertitle{Towards a More Reliable and Available \docker-based Container 
Cloud}

\input{preamble}

\begin{document}

\title{\Large \bf \papertitle}

%for single author (just remove % characters)
\author{
{\rm Mudit Verma}\\
IBM Research
\and
{\rm Mohan Dhawan}\\
IBM Research
} % end author

\maketitle

\begin{abstract}
\small

Operating System-level virtualization technology, or \textit{containers} as they 
are commonly known, represents the next generation of light-weight 
virtualization, and is primarily represented by \docker{}. However, \docker{}'s 
current design does not complement the SLAs from \docker-based container cloud 
offerings promising both reliability and high availability. The tight 
coupling between the containers and the \docker\ daemon proves fatal for the 
containers' uptime during daemon's unavailability due to either failure or 
upgrade.
We present the design and implementation of \tool, which fundamentally isolates 
the containers from the running daemon. Our evaluation shows that \tool\ imposes 
only moderate overheads even under load, while achieving much higher container 
availability.

\end{abstract}

\input{sections/introduction.tex}
\input{sections/overview.tex}
\input{sections/implementation.tex}
\input{sections/evaluation.tex}
\input{sections/relatedwork.tex}
\input{sections/conclusion.tex}

\clearpage

\raggedright
\small

\bibliographystyle{acm}
\bibliography{paper}

% \theendnotes

%%\justifying
%%\appendix
%%\input{sections/appendix}

\end{document}

%% file: preamble.tex
\usepackage{color}

\pdfoutput=1

\definecolor{linkcol}{rgb}{0,0,1}
\definecolor{citecol}{rgb}{0,0.5,0}
\definecolor{urlcol}{rgb}{0.3,0,0}

\usepackage{tikz}
\usepackage{amsmath}
\usepackage{ragged2e}
\usepackage{cite}
\usepackage{txfonts}
\usepackage{fancyhdr}
\usepackage{amssymb}
\usepackage{fancyvrb}
\usepackage{graphicx}
\usepackage{times}
\usepackage{pifont}
\usepackage[hyphens]{url}
\usepackage{xspace}
\usepackage{epstopdf}

\usepackage[aboveskip=5pt,small,labelfont=bf]{caption}
\usepackage{subcaption}

\DeclareCaptionType{copyrightbox}

\usepackage{sty/algorithm2e}
\usepackage{sty/multirow}
\usepackage{sty/flushend}
\usepackage{sty/usenix}
\usepackage{sty/epsfig}
\usepackage{sty/endnotes}

\SetAlFnt{\small}
\SetAlCapFnt{\small}
\SetAlCapNameFnt{\small}
\makeatletter
\def\url@myurlstyle{%
   \@ifundefined{selectfont}{\def\UrlFont{\small}}{\def\UrlFont{\small}}}
   \makeatother
\def\UrlFont{\small\ttfamily}
\newcommand\xref[1]{\S~\ref{#1}}

\newcounter{myctr}
\newenvironment{mylist}{\begin{list}{(\textbf{\arabic{myctr}})}
{\usecounter{myctr}
\setlength{\topsep}{0.5mm}\setlength{\itemsep}{0.25mm}
\setlength{\parsep}{0.25mm}
\setlength{\itemindent}{0mm}\setlength{\partopsep}{0mm}
\setlength{\labelwidth}{-2mm}
\setlength{\leftmargin}{0mm}}}{\end{list}}

\newcounter{myctr1}

\newenvironment{mybullet}{\begin{list}{$\bullet$}
{\setlength{\topsep}{0.5mm}\setlength{\itemsep}{0.25mm}
\setlength{\parsep}{0.25mm}
\setlength{\itemindent}{0mm}\setlength{\partopsep}{0mm}
\setlength{\labelwidth}{-2mm}
\setlength{\leftmargin}{0mm}}}{\end{list}}

\newcommand{\myparagraph}[1]{\noindent{\scshape \bfseries #1.}}

\newcommand{\code}[1]{\texttt{\small{#1}}}
\newcommand{\codesm}[1]{\texttt{\scriptsize{#1}}}
\newcommand{\mycomment}[1]{}

\newcommand{\ie}{{i.e.,}}

\newcommand{\docker}{Docker}
\newcommand{\ltool}{{$\ell$-\tool}\xspace}

\SetEndCharOfAlgoLine{}

\usepackage{hyperref}

%% file: sections/introduction.tex
\section{Introduction}
\label{sec:introduction}

%% docker, docker based container clouds

Container or OS-based virtualization is becoming increasingly popular to provide 
isolation for applications by leveraging the same underlying OS 
kernel~\cite{price:lisa:2004, soltesz:eurosys:2007, bhattiprolu:siops:2008}. 
While several commercial container implementations~\cite{docker, rkt, lxd, 
runc} exist, \docker{}~\cite{docker} provides an easy to use application 
packaging and distribution mechanism, resulting in increased popularity, and 
several \docker{}-based container cloud offerings from Google~\cite{kubernetes}, 
IBM~\cite{bluemix}, Microsoft~\cite{hyperv}, Joyent~\cite{joyent}, amongst 
others.

\begin{figure*}[t!]
\centering
\begin{tabular}{ccc}

\begin{minipage}[b]{0.3\linewidth}
\centering%
\includegraphics[bb=1 121 543 443,height=.725\linewidth]{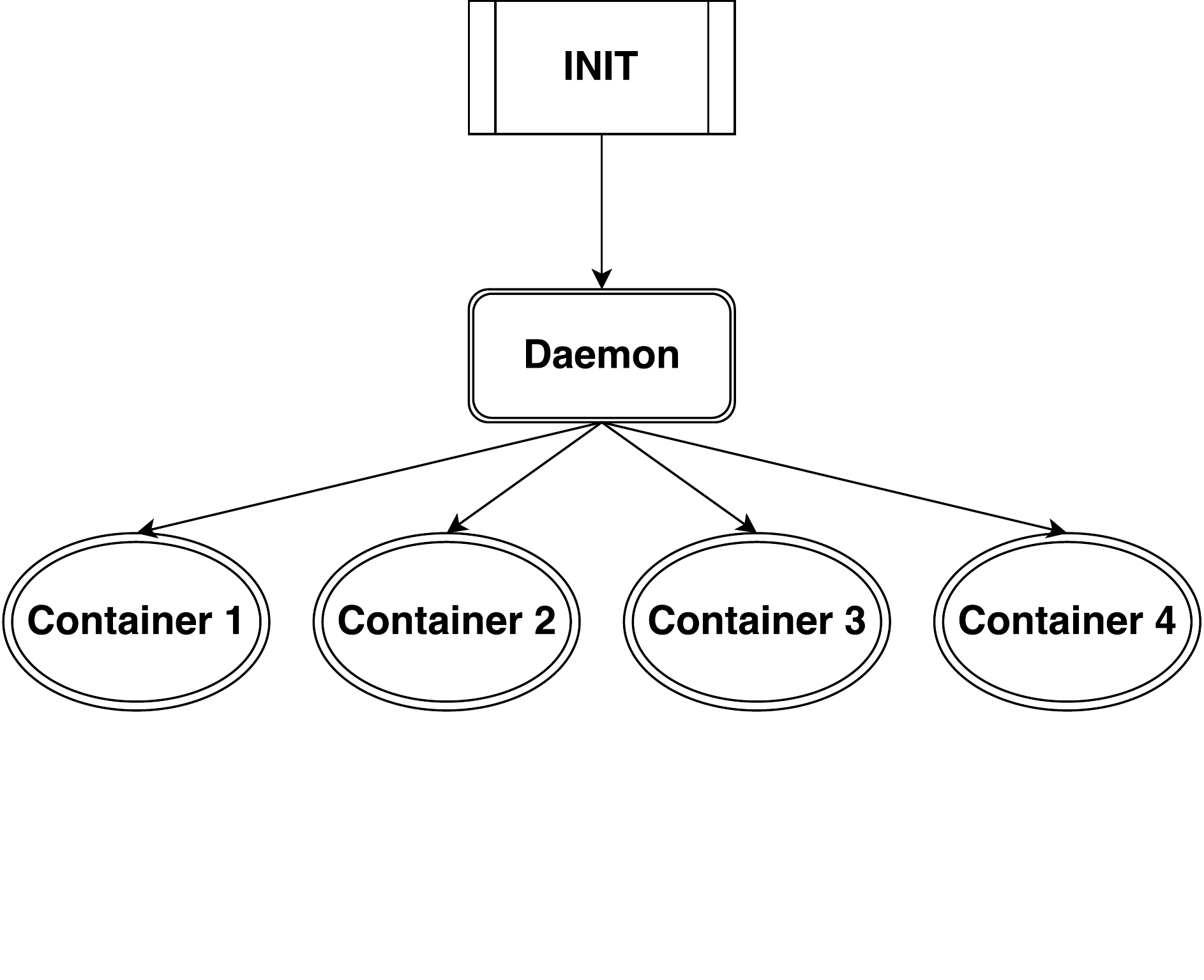}
\caption{\docker\ process tree.}
\label{figure:docker}
\end{minipage}

&

\hspace{-10pt}

\begin{minipage}[b]{0.3\linewidth}
\centering%
\includegraphics[bb=-1 120 681 443,height=.725\linewidth]{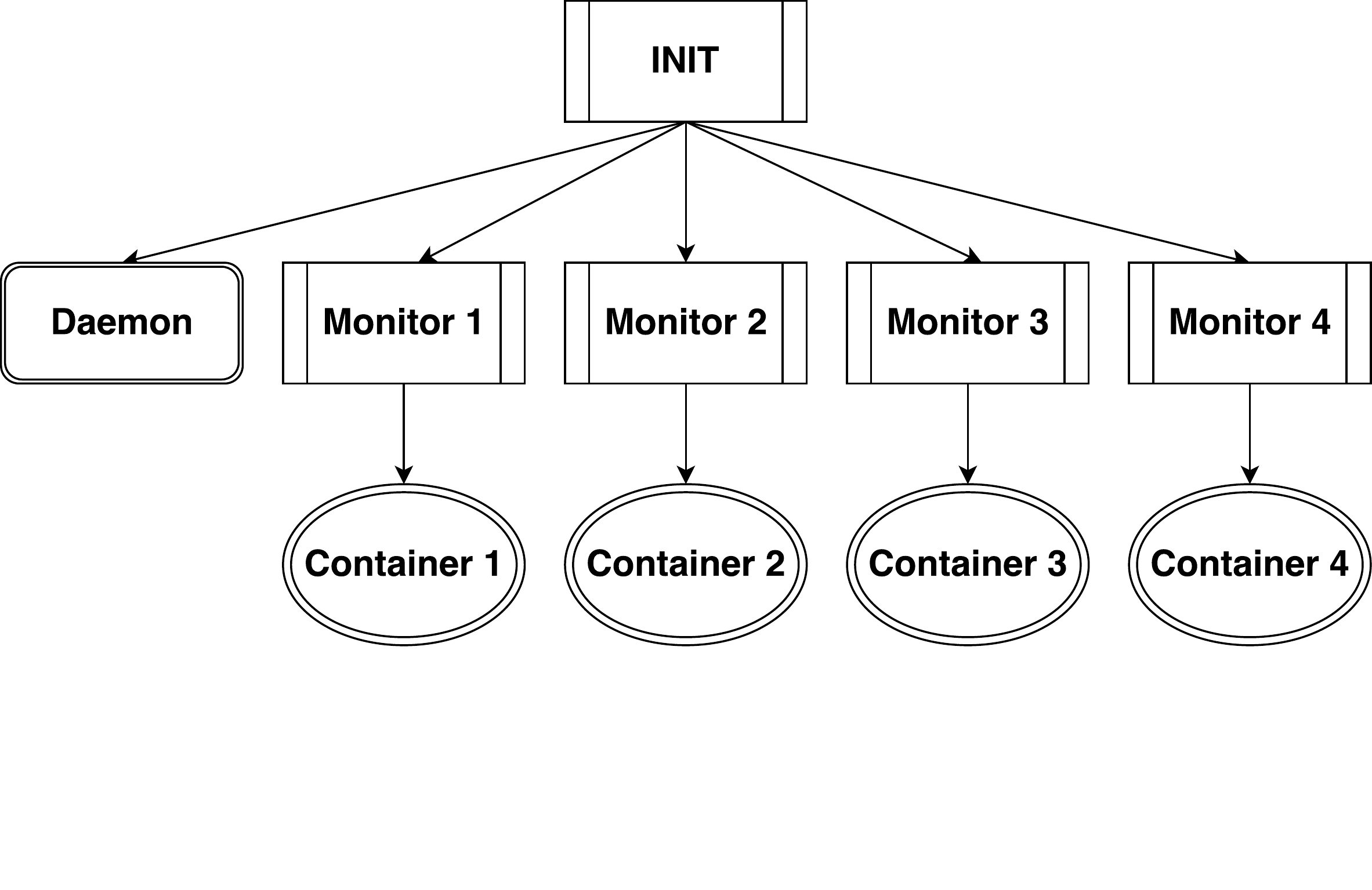}
\caption{\tool\ process tree.}
\label{figure:tool}
\end{minipage}

&

\hspace{15pt}

\begin{minipage}[b]{0.3\linewidth}
\centering%
\includegraphics[bb=-1 0 551 443,height=.725\linewidth]{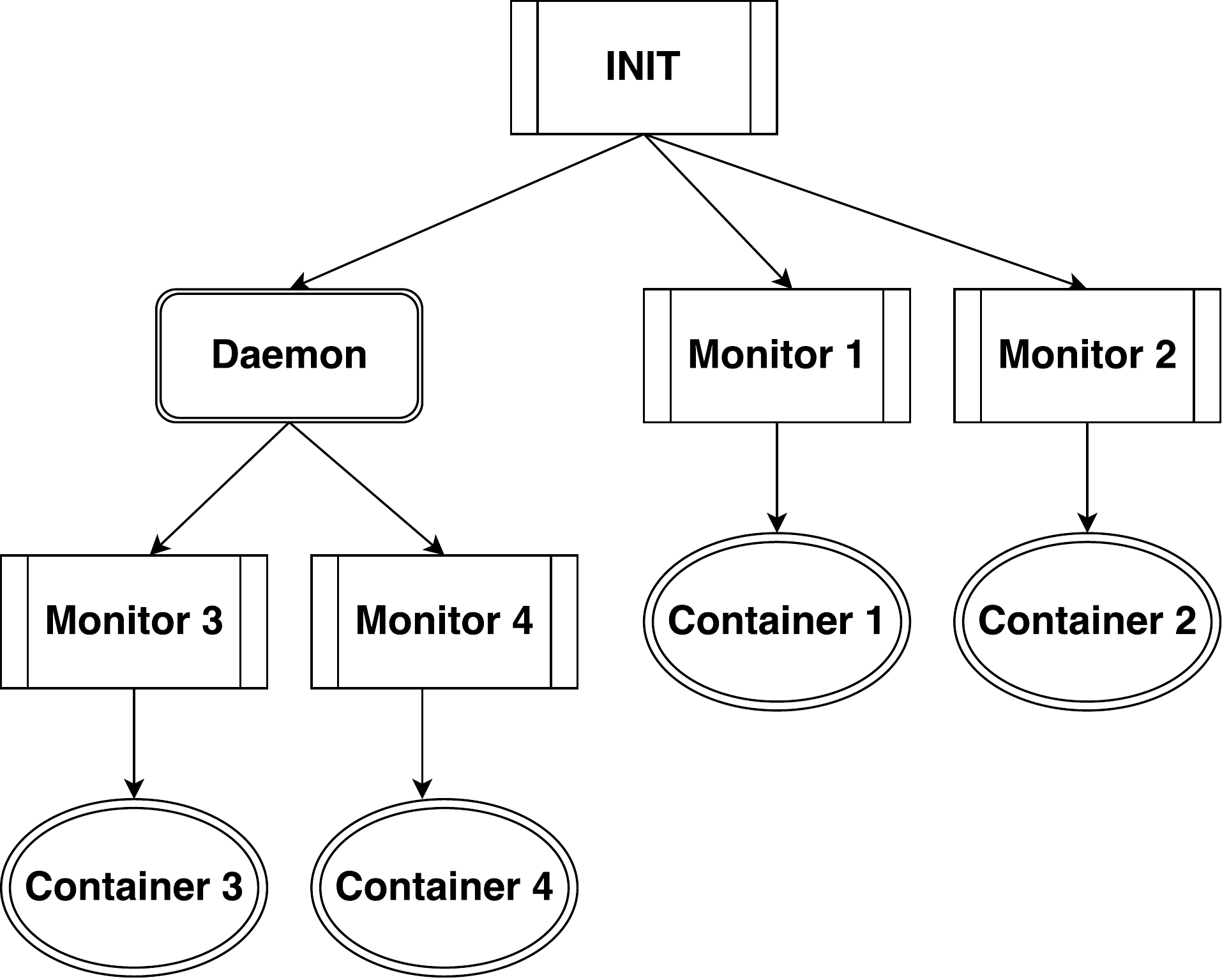}
\caption{\ltool\ process tree.}
\label{figure:lazy-tool}
\end{minipage}

\end{tabular}
% \caption{architectures.}
\end{figure*}

%% what is the problem
While these \docker{}-based cloud services offer strict SLAs promising both 
reliability and availability, just like VM-based clouds, the existing \docker\ 
architecture itself becomes a hindrance. Specifically, the spawned containers 
are direct descendants of the \docker\ daemon (see Fig.~\ref{figure:docker}), 
which is responsible for container creation, state management, and monitoring 
and communication with the containers. This tight coupling between the daemon 
and the containers makes the daemon a single point of failure, thereby causing 
the daemon's unavailability
% and performance 
to affect the containers' uptime.
% and their performance.
Although, containers (along with their applications) do not require the 
daemon's presence for their functioning, \docker\ still aborts all containers 
upon a daemon restart, crash or upgrade.

%% give real world examples
%% any prior work in this regard
Reliability of the \docker\ daemon has been discussed recently~\cite{issue-1, 
issue-2, issue-3} in the context of container clouds. Kubernetes reported 
several \docker\ daemon crashes due to \docker\ bugs, out-of-memory/disk errors, 
corrupt \docker\ pulls, etc., immediately aborting all containers. In IBM's 
container cloud, all running containers must be aborted upon every \docker\ 
upgrade, which occurs frequently.

%% why is it important
The above problem is an example of how a single point of failure impacts the 
entire system. While it has been solved in other contexts, the problem is 
relevant for today's \docker{}-based container cloud, where different containers 
running on a server might belong to same or different clients and might host 
highly available or stateless services. In such scenarios, a container downtime 
caused by external factors such as a daemon's death/upgrade, is highly 
undesirable.
% Specifically, there are three main concerns:
% 
% \begin{mylist}
%  \item In its present form, container availability is strictly tied to the 
% \docker\ daemon. For example, restarting the \docker\ daemon in event of a crash 
% due to bugs in \docker\ code or software upgrades instantly aborts all 
% associated containers. Large servers hosting hundreds of containers, often 
% belonging to different clients, cannot afford to have downtime due to a crashed 
% daemon or a hot upgrade.
% 
%  \item Redundancy amongst the containers does not improve their reliability in 
% case of daemon failure, unless these redundant containers execute and bind to 
% \textit{another} \docker\ daemon, preferably running on a separate node. 
% However, such an arrangement would require significant application-level 
% book-keeping to maintain availability of the service running within the 
% containers.
% 
%  \item Containers are likely to suffer from control plane performance issues 
% introduced by the \docker\ daemon, since it is a heavy weight process. For 
% example, since containers and the daemon execute atop the same kernel, a daemon 
% utilizing inordinate amounts of CPU or memory resources, would significantly 
% affect container execution.
% \end{mylist}

%% our solution
We present the design~(\xref{sec:tool}) and 
implementation~(\xref{sec:implementation}) of \tool, which decouples the 
\docker\ daemon and the containers without affecting any functionality. 
Specifically, \tool\ re-organizes the process dependencies amongst the daemon 
and containers such that the running containers and daemon are siblings in 
process tree, and are not affected even in case of daemon's unavailability. 

% When the client request creation of a container, the daemon first spawns a 
% light-weight monitor process, which then subsequently spawns the actual 
% container. \tool\ then \textit{daemonises} the monitor process so that monitor 
% (along with the container) become orphans and are adopted by the top-level INIT 
% process. This mechanism detaches the containers from the daemon at the instant 
% they are spawned. The \docker\ daemon interacts with containers via (a) the 
% monitor process using sockets, or (b) direct signals to container process. Once 
% a container finishes execution, its monitor process communicates the container's 
% exit status to the daemon. 

We have built a prototype of \tool, and our evaluation~(\xref{sec:evaluation}) 
shows that \tool\ allows upgrades to the daemon without aborting any 
containers. Furthermore, with $500$ Redis and $200$ Tomcat containers, \tool\ 
improves launch latencies by $\sim$$53\%$ (or $800$ ms) and $\sim$$33\%$ (or 
$190$ ms), respectively, over vanilla \docker. Additionally, while \docker\ 
daemon's resident memory utilization increases linearly with the number of 
containers, \tool\ daemon's memory usage increases by just $10$ MB up to $20$ 
MB even for $500$ Redis containers, \ie\ a saving of $\sim$$92\%$ over vanilla 
\docker.

%% file: sections/overview.tex
\section{\tool}
\label{sec:tool}

% The availability of \docker\ daemon and the containers is highly susceptible 
% to  bugs in \docker\ code, accidental process crash, imperfect software 
% upgrades, etc. 

We identify broad design goals for a highly reliable and available container 
architecture as follows:

\begin{mylist}
 \item \textbf{Minimal container downtime}: Daemon unavailability must not 
affect a container's uptime, and container applications must continue 
functioning as normal.

 \item \textbf{High daemon availability}: The daemon must have minimal downtime. 
In the event of a crash, it must recover its previous state quickly.

 \item \textbf{Low overhead}: Improvements in reliability and availability must 
not adversely impact the control plane operations between the daemon and the 
containers.
\end{mylist}

Lastly, while it would be desirable to seamlessly retrofit existing \docker\ 
deployments with the above properties, we do not make it a focus of this work.

\subsection{Design}
\label{sec:tool:design}

\myparagraph{Key Idea} \tool\ re-architects the process dependencies to 
eliminate strong coupling between containers and the daemon. Specifically, 
\tool\ leverages the observation that an orphaned Linux process is adopted 
naturally by the top-level INIT process. Thus, \tool\ decouples a container 
process from the daemon just after the container's first process is spawned, 
such that the container and the daemon become siblings in the process hierarchy.

In \tool, on receiving a client request to create a container, the daemon first 
spawns a per-container \textit{monitor} process, which subsequently spawns the 
container. Once the container process starts, the monitor replaces itself (using 
UNIX \code{exec}) with an ultra-lightweight code, whose primary responsibility 
is to assist in communication between the container and the daemon. \tool\ then 
\textit{daemonises} the monitor, \ie\ the parent-child relationship between the 
daemon and the monitor is decoupled, following which the orphaned monitor (along 
with its container) is adopted by the top-level INIT. Fig.~\ref{figure:tool} 
shows the process hierarchy for \tool.

\subsubsection{Per-container monitor process}
\label{sec:tool:design:monitor}

\tool's per-container monitor is stateless. However, it is responsible for 
communicating container's state to the daemon, and managing its interaction with 
the \docker\ client. The monitor must communicate the container's (a) \code{pid} 
(when the container is spawned), and (b) exit status (when the container 
finishes execution), for state management within the daemon. In order to 
reproduce all \docker\ functionality, the monitor must enable mechanisms to 
attach and re-direct I/O streams to the client.

% There are two possibilities here: Daemon is alive:- communication is straight 
% forward and the monitoring process can exit after supplying the status to the 
% daemon. Daemon is dead:- Monitoring process will wait until the daemon is 
% up, communicate the status and exit. Alternatively it can write to a file, 
% which the new instance of daemon when it would come back up, will read from. 

\subsubsection{Daemon -- container communication}
\label{sec:tool:design:communication}

\docker\ currently leverages the parent-child hierarchy for communication 
between the daemon and containers. For example, the daemon performs a 
\code{wait} on the container process in a separate thread to determine its 
execution status. However, with the restructuring of this process hierarchy in 
\tool, these control plane communication channels between the daemon and the 
containers no longer exist. In order to retain the existing \docker\ 
functionality, \tool\ facilitates all control plane operations leveraging 
out-of-band communication mechanisms between the daemon and the per-container 
monitor process.
% We list below some salient features of this bi-directional communication in 
% \tool.

\begin{mylist}
 \item \textbf{Container to daemon}: The container to daemon communication is 
mediated by the monitor process. Once the container terminates, the waiting 
monitor process in \tool\ (i) writes the exit status in a file, (ii) sends a 
special signal (like \textsf{\scriptsize SIGRTMIN}-\textsf{\scriptsize 
SIGRTMAX}, \textsf{\scriptsize SIGUSR1}, \textsf{\scriptsize SIGUSR2}) to the 
daemon to read the file, and (iii) exits itself. If the daemon is unavailable 
or dead, then upon a subsequent restart, the daemon reads the saved state from 
the file and performs the necessary state management.

 \item \textbf{Daemon to container}: The daemon communicates with the container 
through direct signaling. The per-container monitor process informs the daemon 
of the container's \code{pid} at its launch, following which the daemon can send 
management instructions to the container (\ie\ stop, pause, kill, continue) 
using signals like \textsf{\scriptsize SIGTERM}, \textsf{\scriptsize SIGSTP}, 
\textsf{\scriptsize SIGKILL}, \textsf{\scriptsize SIGCONT}, etc.

 \item \textbf{Control plane I/O}: \tool\ leverages interprocess pipes to 
redirect all I/O streams (\code{stdin}, \code{stdout} and \code{stderr}) 
between the \docker\ client and the containers.
\end{mylist}

\subsection{Availability and Reliability}
\label{sec:tool:design:reliability}

In event of a daemon restart, following a crash or upgrade, \docker\ reboots all 
previously executing containers associated with the daemon.
% , and subsequently kills them to clean all corresponding state. 
\tool, however, does not terminate any running container, since the containers 
are not tied to a particular daemon instance. Thus, \tool, by design,
% significantly reduces any 
avoids container downtime due to daemon unavailability.

Decoupling the parent-child association of the daemon and the containers allows 
per-container monitor processes to communicate with \textit{any} instance of the 
daemon, since, unlike existing \docker\ implementations, no particular daemon 
\textit{owns} the container in \tool. Thus, in principle, this distributed 
association will help \tool\ achieve high availability for the daemon itself, by 
scaling up the daemon instances on the same/different host(s) and performing 
load balancing on the distributed requests from the CLI or remote APIs.

\myparagraph{Recovering from a monitor crash} \tool\ does not \textit{eliminate} 
the possibility of container downtime. While the monitor process may itself be 
terminated inadvertently, it would affect just a single container.
% 's uptime. 
% Thus, \tool\ merely reduces the impact of daemon (or monitor) unavailability on 
% a container's lifetime.
In the event of a monitor crash, the orphaned container is adopted by the INIT, 
and continues to function normally without any downtime. However, the daemon has 
no way to control/interact with such a container. \tool's daemon overcomes 
this problem by (a) maintaining a list of \code{pid} for both the monitor and 
the container, and (b) periodic polling for the monitor processes in a 
dedicated thread. When the daemon detects a live but orphaned container (based 
on the \code{pid} associations), it forcefully takes down and reboots the 
container, thereby maintaining connectivity with the monitor and the container.

% Even though the monitor has minimal code and exercises minimal functionality, it 
% might inadvertently be killed.

% \subsection{Benefits}
% \label{sec:tool:design:benefits}
% 
% \tool's re-structuring of the process hierarchy enables high availability for 
% the containers by design, thereby allowing hot upgrades or daemon restarts 
% without container downtime. Furthermore, decoupling the parent-child 
% association of the daemon and the containers also allows per-container monitor 
% processes to bind to \textit{any} instance of the daemon, since, unlike 
% existing \docker\ implementations, no particular daemon \textit{owns} the 
% container in \tool. This distributed association helps \tool\ achieve high 
% availability for the daemon itself, by scaling up the daemon instances on the 
% same host and performing load balancing on the distributed requests from the 
% CLI or remote APIs.

\subsection{Lazy-\tool (\ltool)}
\label{sec:tool:design:ltool}

In \textit{lazy} \tool{}, the newly spawned monitors (along with their 
containers) are initially direct children of the daemon, as in current \docker\ 
implementation. However, upon daemon failure (due to crash or upgrade), all the 
monitors (along with their running containers) become orphan and are adopted 
naturally by the top-level INIT process. Fig.~\ref{figure:lazy-tool} presents 
the process tree for \ltool. While equivalent to \tool\ in functionality, 
\ltool's design has less homogeneity, with currently spawned monitors (along 
with containers $1$ and $2$) attached as descendants of the current daemon 
instance, and orphaned monitors along with containers $3$ and $4$ (due to a 
previous daemon crash) attached to the INIT process.

%% file: sections/implementation.tex
\section{Implementation}
\label{sec:implementation}

We enhance vanilla \docker\ v$1.8.0$  to build 
a prototype of \tool\ (per~\xref{sec:tool:design}). We below describe key 
features of our implementation.

\begin{mybullet}
 \item \textbf{Monitor creation}: On every new container spawn, \tool\ daemon 
double forks the \docker\ binary. The last forked \docker\ process further 
spawns the container process as its descendant and subsequently replaces itself 
(using UNIX \code{exec}) with a light weight monitor code.

 \item \textbf{Process hierarchy re-organization}: Once the monitor creation is 
complete, the monitor kills its parent, \ie\ a forked intermediate process, 
using \code{ppid} and UNIX \code{kill}, thereby \textit{daemonising} itself. The 
monitor (along with its container) is then adopted by INIT, thereby completing 
the re-organization of the process hierarchy.
 
 \item \textbf{Out-of-band communication}: The monitor implements a \code{wait} 
on its descendant container process in a thread, and supplies the container's 
exit status to the daemon by (a) writing to a specific file (named by the 
containerId) at a pre-defined path in the container's workspace, and (b) 
sending a special signal \textsf{\scriptsize SIGRTMIN+10} from the reserved 
signal pool \textsf{\scriptsize SIGRTMIN}-\textsf{\scriptsize SIGTMAX} to the 
daemon, which upon receiving this signal, reads the exit status from the 
designated file.
% 's pid using \code{kill(pid, signal)}. 
In another thread, the monitor implements an API server, which listens to its 
own unique socket (containerId.sock). The daemon maintains and supplies this 
socket information to the \docker\ client, following which the client connects 
directly to the monitor for container I/0 redirection.
\end{mybullet}

\tool\ required modifications to $\sim$$1300$ LOC across $25$ files, impacting 
just $12$ of the total $40$ \docker\ commands~\footnote{\scriptsize The affected 
\docker\ commands are \codesm{attach}, \codesm{exec}, \codesm{kill}, 
\codesm{pause}, \codesm{restart}, \codesm{run}, \codesm{start}, \codesm{stats}, 
\codesm{stop}, \codesm{top}, \codesm{unpause}, and \codesm{wait}.}. The above 
changes represent $<$$1\%$ of \docker{}'s code base in terms of LOC, and 
$<$$3\%$ in terms of the total files. Thus, only minimal changes are required to 
significantly enhance the reliability and availability of the containers. We 
also implemented \ltool\ as described in~\xref{sec:tool:design}. \ltool\ 
leverages most of \tool's code base except that it requires single fork (instead 
of a double fork described above), and also does not need subsequent 
\textit{daemonising} of the monitor.

%% file: sections/evaluation.tex
\begin{figure*}[t]
\centering
\begin{tabular}{ccc}

\begin{minipage}[b]{0.3\linewidth}
\centering%
\includegraphics[width=\linewidth]{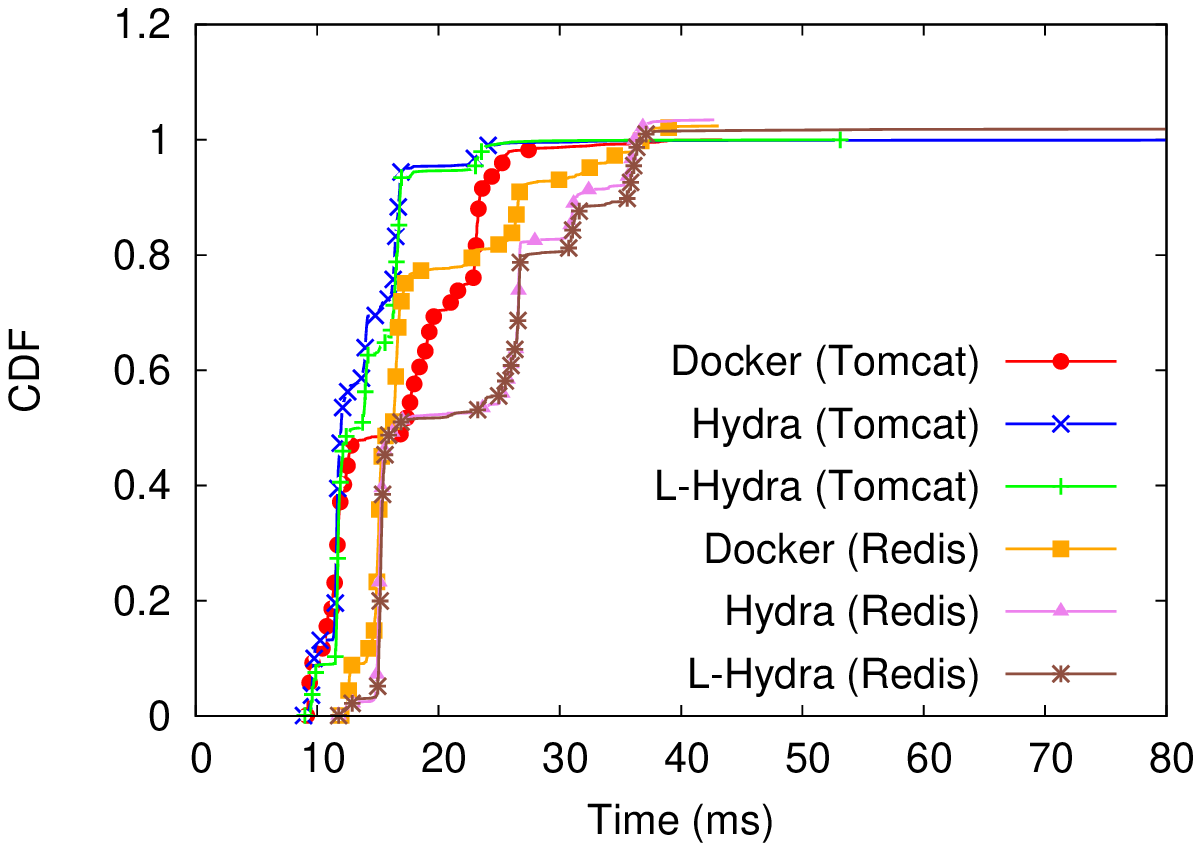}
\subcaption{\scriptsize Time to restore daemon.}
\label{figure:daemon-restart}
\end{minipage}

&

\begin{minipage}[b]{0.3\linewidth}
\centering%
\includegraphics[width=\linewidth]{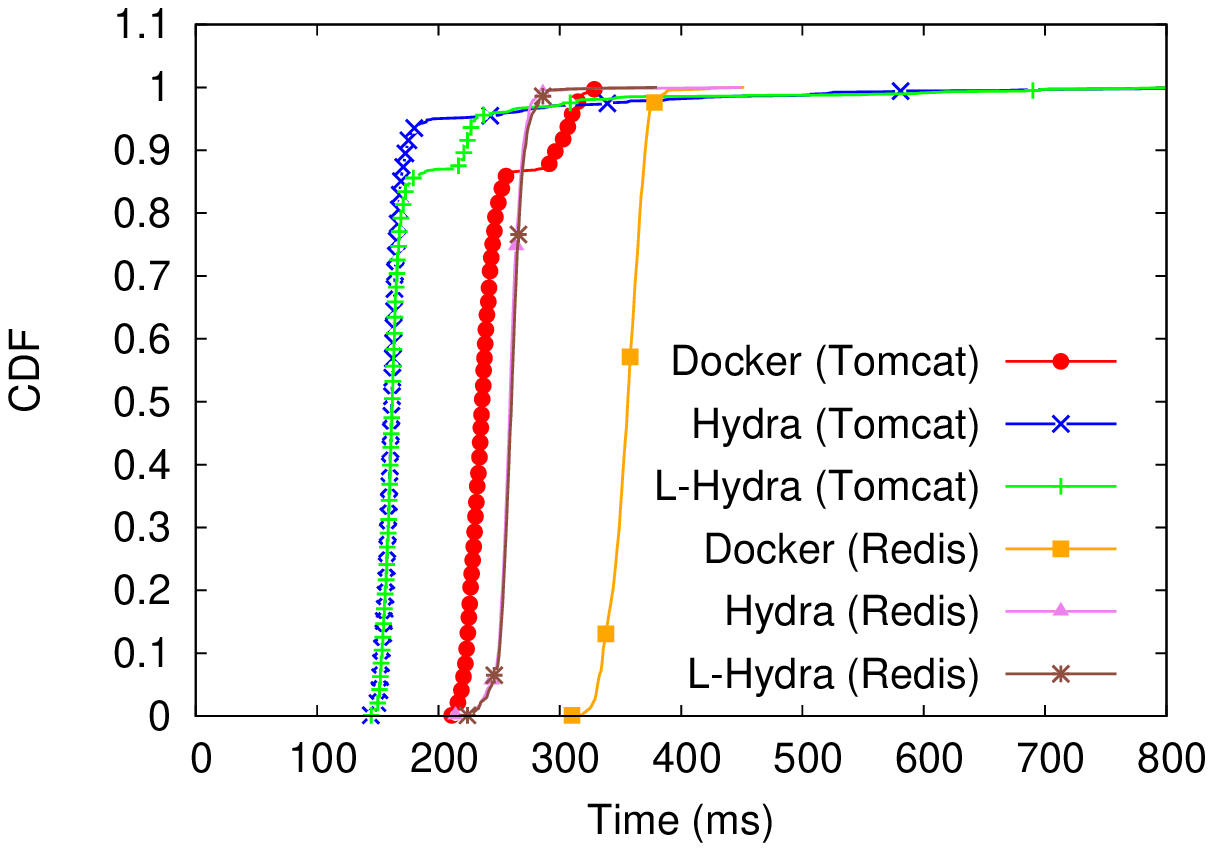}
\subcaption{\scriptsize Time to restore running container.}
\label{figure:container-restart}
\end{minipage}

&

\begin{minipage}[b]{0.3\linewidth}
\centering%
\includegraphics[width=\linewidth]{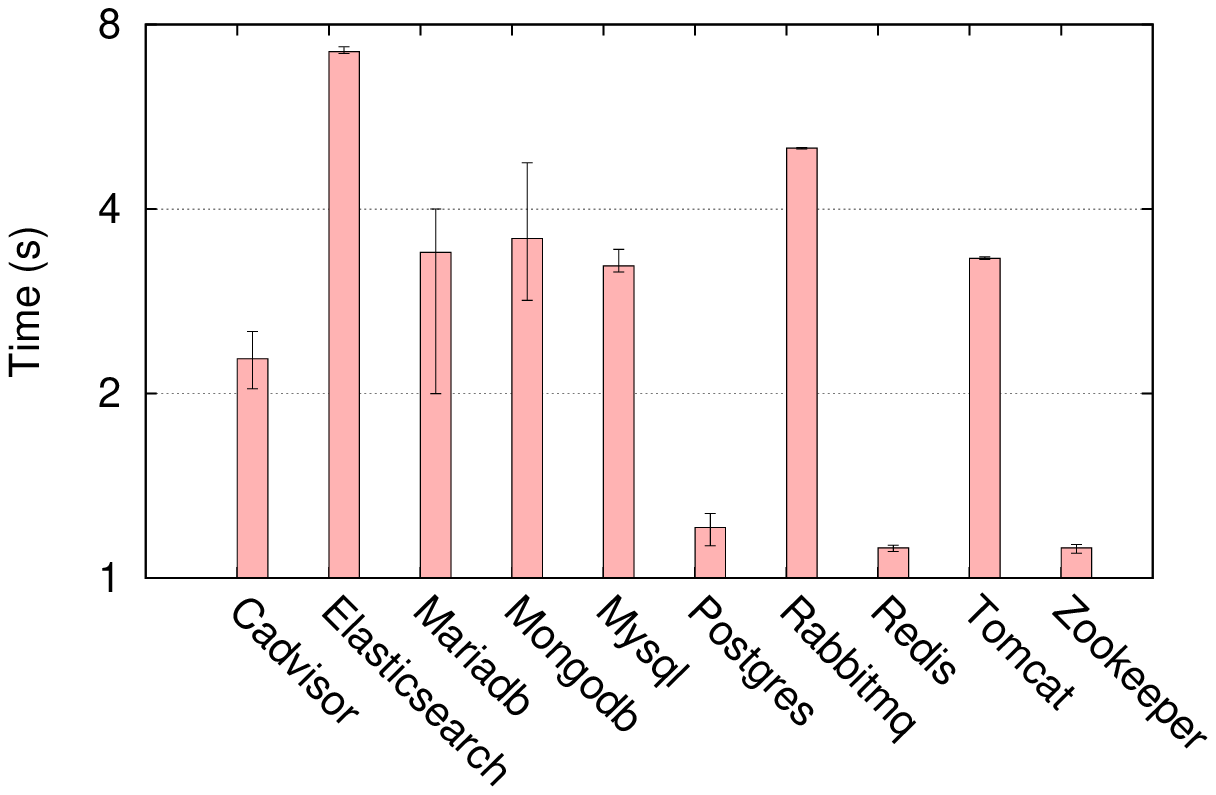}
\subcaption{\scriptsize Application outage time for vanilla \docker.}
\label{figure:app-outage}
\end{minipage}

\end{tabular}
\caption{Reliability comparison of \tool\ and \ltool\ against vanilla \docker.}
\end{figure*}

\section{Evaluation}
\label{sec:evaluation}

In~\xref{sec:evaluation:reliability}, we demonstrate the reliability provided by 
\tool\ under varying conditions, and quantify the impact of a daemon crash on 
daemon, container and application. In~\xref{sec:evaluation:performance}, 
we measure perceived end user latencies for \tool. Lastly, 
in~\xref{sec:evaluation:scalability}, we compare scalability of \tool\ against 
vanilla \docker.

\myparagraph{Experimental Setup} All experiments were performed on a virtual 
machine provisioned with $16$ cores at $2.0$ GHz, $32$ GB RAM, and running $64$ 
bit Ubuntu v$14.04$. We compare \tool\ and \ltool\ against vanilla \docker\ 
v$1.8.0$. Unless stated, we evaluate \ltool\ 
considering a daemon that has not abruptly crashed, \ie\ new containers attach 
to the daemon's process hierarchy. Further, all experiments 
(except~\xref{sec:evaluation:scalability}) assume no previously running 
containers, and use official, unmodified container images.

\subsection{Reliability}
\label{sec:evaluation:reliability}

A daemon restart entails (i) restoring daemon state, (ii) rebooting all 
previously running containers, and subsequently (iii) initializing container 
applications to make them operational. In the experiments below, we first launch 
an application container and subsequently kill the daemon process. We then 
compute savings for both \tool\ and \ltool\ over vanilla \docker\ on a daemon 
restart by measuring time to restore the daemon and containers, and time to 
complete application initialization (after container has been restored). We also 
measure the perceived application outage due to \docker\ upgrade.
% , \ie\ time between daemon shutdown and subsequent application state 
% restoration.

% \begin{mylist}
\myparagraph{(i) Daemon restart} Redis is a light-weight application that runs 
few processes and has low initialization time, while Tomcat is a heavy-weight 
application with several processes and high initialization time. Thus, both 
Redis and Tomcat lie at opposite ends of the application  spectrum. We repeat 
each experiment $1K$ times with just one running container. Further, default log 
levels, and container and application configurations were enabled.
% \myparagraph{(i) State restoration} We consider Redis and Tomcat applications 
% for evaluation, and repeat each experiment $1K$ times with just one running 
% container, and default log levels enabled.

\begin{mybullet}
 \item \textbf{Daemon}: Fig.~\ref{figure:daemon-restart} shows a CDF of the 
state restoration time for the daemon upon a restart from a crash. We note that 
across both the applications, the daemon state restoration time remained fairly 
consistent at $<40$ ms for \tool, \ltool\ and vanilla \docker.

 \item \textbf{Container}: Unlike \docker, \tool\ and \ltool\ do not need to 
\textit{restore} previously running containers, since their containers do not 
abort on a daemon crash. Instead, \tool\ and \ltool\ simply read state from 
already running containers, leading to lower container restoration time. 
Fig.~\ref{figure:container-restart} shows the results. We note that across the 
benchmarks, \tool\ and \ltool\ report container restoration times $\sim$$100$ ms 
lower than vanilla \docker.
% Further, the container restoration time for Tomcat was less than Redis for 
% vanilla \docker, possibly due to the light-weight nature of Redis.

\item \textbf{Application}: Since a daemon crash does not impact running 
containers in \tool\ and \ltool, the application startup time is applicable only 
for vanilla \docker\ setup. Note that initialization depends significantly 
on the nature of the application. Our experiments reveal that after a restart, 
Redis completed initialization in $\sim$$2$ ms, while Tomcat required 
$\sim$$1039$ ms. Furthermore, restart times may be lower than that of fresh 
application starts, especially for stateful applications, like MySQL, which 
reloads stored persistent state upon restart. For example, we observed that a 
fresh initialization for MySQL required $\sim$$5600$ ms, while a restart took 
just $\sim$$300$ ms.
\end{mybullet}

\myparagraph{(ii) Outage during upgrades} We select $10$ popular, but diverse, 
official container images (with each having several million downloads), and 
measure the user perceived application outage due to an upgrade from v$1.8.0$ to 
v$1.8.3$ for vanilla \docker. Recall that both \tool\ and \ltool\ do not incur 
any container downtime. We perform each experiment $10$ times, and report the 
average application outage in Fig.~\ref{figure:app-outage}. We observe that 
both \tool\ and \ltool\ save upon application outages of several seconds as 
compared to \docker.
% \end{mylist}

\begin{figure*}[t]
\centering
\begin{tabular}{ccc}

\begin{minipage}[b]{0.3\linewidth}
\centering%
\includegraphics[width=\linewidth]{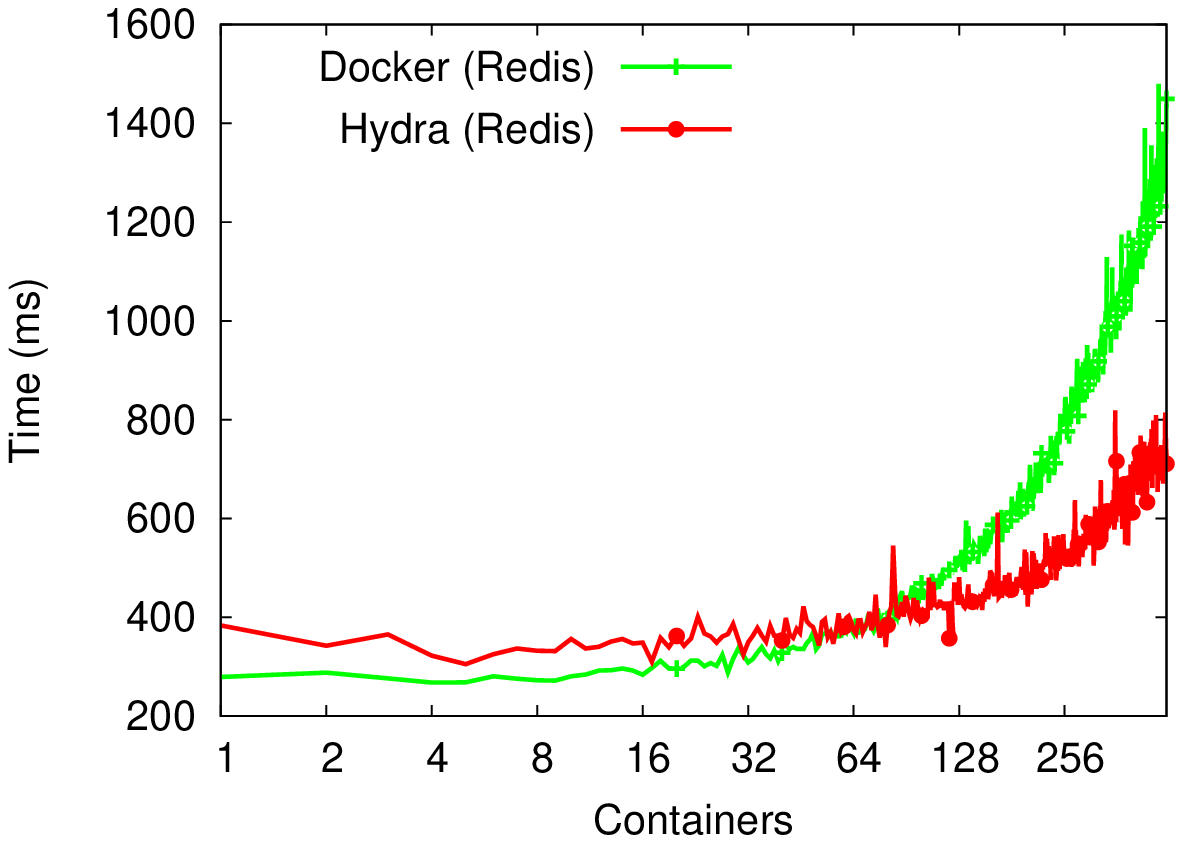}
\subcaption{\scriptsize Redis launch times for \tool\ \& \docker.}
\label{figure:scalability-redis}
\end{minipage}

&

\begin{minipage}[b]{0.3\linewidth}
\centering%
\includegraphics[width=\linewidth]{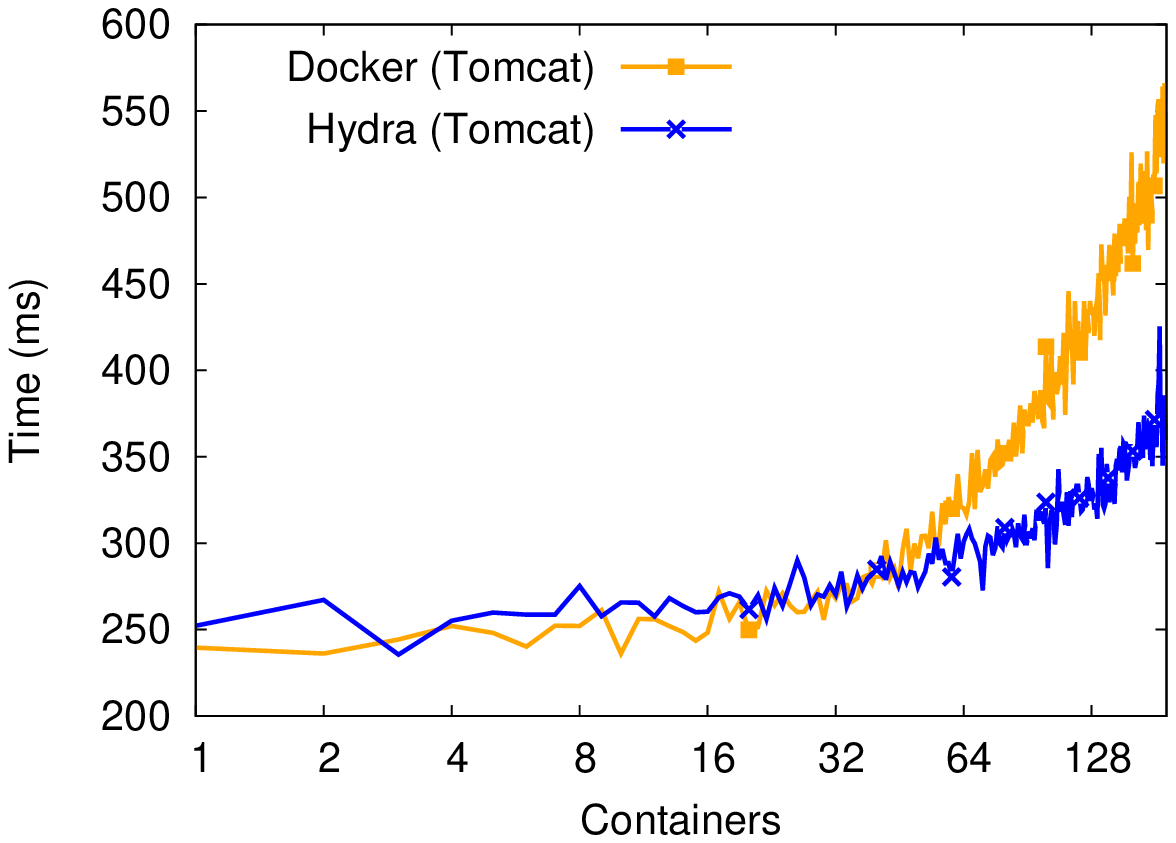}
\subcaption{\scriptsize Tomcat launch times for \tool\ \& \docker.}
\label{figure:scalability-Tomcat}
\end{minipage}

&

\begin{minipage}[b]{0.3\linewidth}
\centering%
\includegraphics[width=\linewidth]{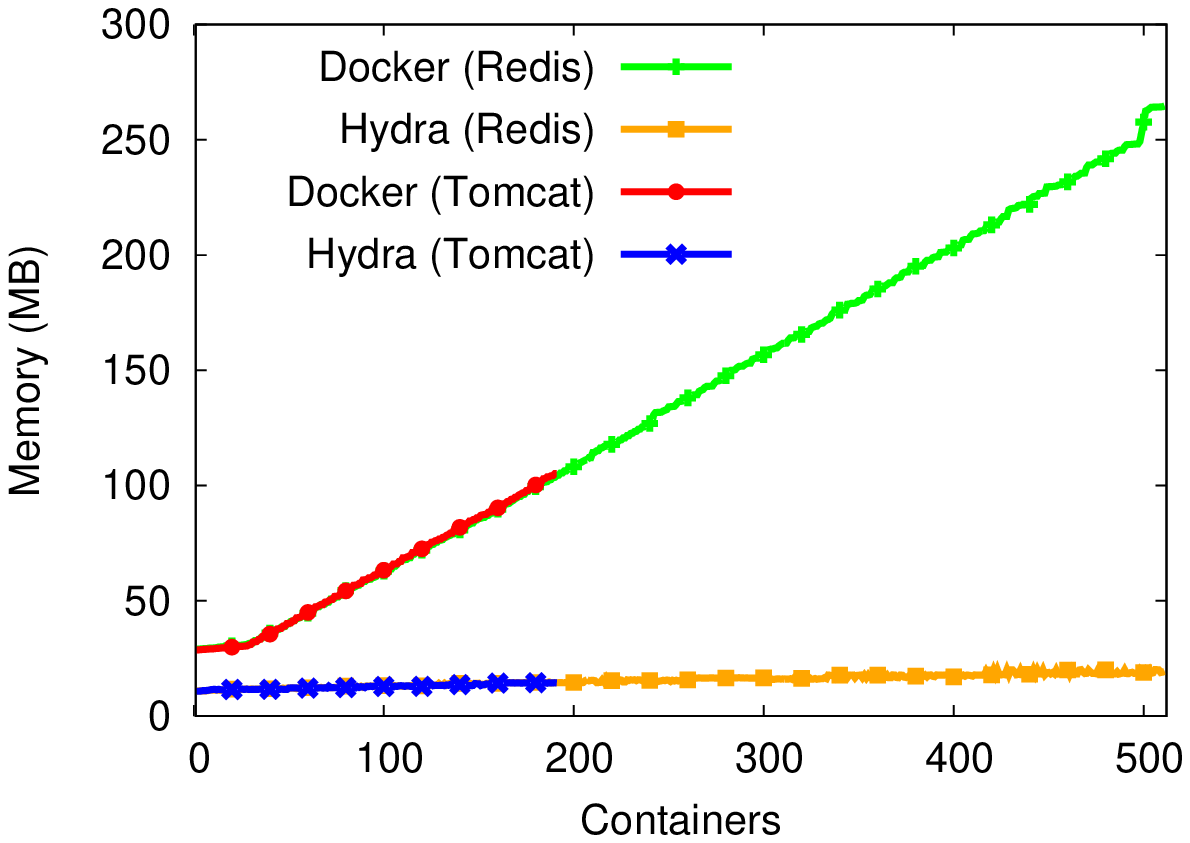}
\subcaption{\scriptsize Daemon memory usage for \tool\ \& \docker.}
\label{figure:scalability-memory}
\end{minipage}

\end{tabular}
\caption{Scalability comparison of \tool\ against vanilla \docker.}
\end{figure*}

\subsection{Performance overheads}
\label{sec:evaluation:performance}

% \begin{mylist}
%  \item \textbf{End-to-end}: 
 
An important measure of performance is the perceived end user latency for 
spawning a container. We use the $10$ previously selected benchmarks (as listed 
in Fig.~\ref{figure:app-outage}), and measure the time to completely spawn a 
container, which includes both the daemon and monitor instantiation, along with 
the application startup time. Note that, 
unlike~\xref{sec:evaluation:reliability}, application startup here corresponds 
to a fresh start with no application specified persistent state, and could thus 
be significantly more than just \textit{restart} times, especially for stateful 
applications, like MySQL.

\begin{table}[t]
\centering
\scriptsize
\setlength{\tabcolsep}{3.5pt}
\begin{tabular}{l|r|r|r|r|r|r}
\multicolumn{1}{c|}{\multirow{2}{*}{\textbf{Image}}} &
\multicolumn{1}{c|}{\textbf{App. Init.}} &
\multicolumn{1}{c|}{\textbf{\docker}} &
\multicolumn{2}{c|}{\textbf{\tool}} &
\multicolumn{2}{c}{\textbf{\ltool}}\\
 & Time (ms) & Time (ms) & Time (ms) & (\%) & Time (ms) & (\%)\\
\hline
Cadvisor & $2256$ & $2503$ & $2637$ & $5.4$ & $2625$ & $4.9$\\
Elasticsearch & $5187$ & $5452$ & $5548$ & $1.8$ & $5527$ & $1.4$\\
Mariadb & $9010$ & $9258$ & $9379$ & $1.3$ & $9348$ & $0.9$\\
Mongodb & $983$ & $1217$ & $1355$ & $11.3$ & $1318$ & $8.3$\\
Mysql & $5621$ & $5848$ & $5975$ & $2.2$ & $5947$ & $1.7$\\
Postgres & $4840$ & $5079$ & $5198$ & $2.3$ & $5171$ & $1.8$\\
Rabbitmq & $3002$ & $3291$ & $3373$ & $2.5$ & $3360$ & $2.1$\\
Redis & $2$ & $228$ & $340$ & $49.1$ & $316$ & $38.6$\\
Tomcat & $1039$ & $1328$ & $1453$ & $9.4$ & $1417$ & $6.7$\\
Zookeeper & $3153$ & $3406$ & $3495$ & $2.6$ & $3463$ & $1.7$\\  
\end{tabular}
\caption{User perceived latencies for starting a container.}
\label{table:end-to-end}
\end{table}

We perform the experiment $10$ times and measure the absolute time and relative 
overheads for every container spawn, for \tool, \ltool, and vanilla \docker. 
Table~\ref{table:end-to-end} reports the average time and overhead for each 
benchmark. We note that most benchmarks, except Mongodb, Redis and Tomcat, 
incur low relative overheads ($\leq$$5\%$) since the application startup time 
masks the overhead due to monitor creation and re-organization of the process 
hierarchy (as described earlier in~\xref{sec:implementation}).

We further breakdown the absolute times for \tool, and observe that the 
latencies due to monitor creation and re-organization of the process hierarchy 
varies between $80$-$140$ ms. Due to the single process fork for monitor 
creation in \ltool, unlike the double fork in \tool, this latency drops 
slightly and ranges between $60$-$120$ ms. Thus, benchmarks with application 
startup times comparable with these latencies, such as Mongodb, Redis and 
Tomcat, incur high relative overheads.

\subsection{Scalability}
\label{sec:evaluation:scalability}

% \begin{table}[t]
% \centering
% \scriptsize
% \setlength{\tabcolsep}{3pt}
% \begin{tabular}{r|r|r|r|r}
% \multicolumn{1}{c|}{\multirow{2}{*}{\textbf{\#Containers}}} &
% \multicolumn{2}{c|}{\textbf{Redis}} &
% \multicolumn{2}{c}{\textbf{MySQL}}\\
%  & \multicolumn{1}{c|}{\textbf{\docker{} (ms)}} &
% \multicolumn{1}{c|}{\textbf{\tool{} (ms)}} &
% \multicolumn{1}{c|}{\textbf{\docker{} (ms)}} &
% \multicolumn{1}{c}{\textbf{\tool{} (ms)}}\\
% \hline
% $1$ & $x$ & $x$ & $x$ & $x$\\
% $10$ & $x$ & $x$ & $x$ & $x$\\
% $100$ & $x$ & $x$ & $x$ & $x$\\
% $1000$ & $x$ & $x$ & $x$ & $x$\\
% $10000$ & $x$ & $x$ & $x$ & $x$\\
% \end{tabular}
% \caption{\tool's impact on scalability.}
% \label{table:scalability}
% \end{table}

We determine \tool's impact on scalability by measuring the time taken to 
launch an $(n$$+$$1)$$^{th}$ Redis or Tomcat container, with $n$ stable 
containers already executing.
% , when scaling the number of application containers exponentially.
% Additionally, we consider Redis and Tomcat for evaluating the impact of 
% different applications on \tool's scalability, \ie\ light-weight against 
% heavy-weight. 
Fig.~\ref{figure:scalability-redis} and~\ref{figure:scalability-Tomcat} show 
a log-scale variation in latencies to launch subsequent Redis and Tomcat 
containers, for both \tool\ and vanilla \docker. Note that we could not spawn 
more that $200$ Tomcat containers on our VM, due to their high memory 
requirements. Redis posed no such issues, and we easily spawned $500$ 
containers.

We observe that for vanilla \docker\ the launch latencies increase linearly as 
the number of containers increase. In contrast, even though \tool\ incurs higher 
container launch overheads initially, it significantly reduces these overheads 
as we scale. Specifically, with $500$$^{th}$ Redis and $200$$^{th}$ Tomcat 
container spawn, \tool\ improves launch latencies by $\sim$$53\%$ (or $800$ ms) 
and $\sim$$33\%$ (or $190$ ms), respectively, over vanilla \docker. When 
operating at scale, these savings in \tool\ primarily stem from not having (a) 
\code{wait} threads for each container within the daemon, and (b) a bloated 
daemon process hierarchy (as in vanilla \docker{}) that significantly increases 
subsequent container fork latencies~\cite{fork}.

Additionally, while \docker\ daemon's resident memory utilization increases 
linearly with the number of containers, \tool\ daemon's memory usage increases 
by just $10$ MB up to $20$ MB even for $500$ Redis containers, \ie\ a saving of 
$\sim$$92\%$ over vanilla \docker. This minor increase in memory usage is due 
to the container state accumulation in the \tool\ daemon. In contrast, \docker\ 
daemon manages several \code{GO}~\cite{go} runtime threads, including a 
\code{wait} thread, for every container launched. Each of these additional 
threads accumulate memory and contribute to the high resident memory usage of 
the \docker\ daemon. Fig.~\ref{figure:scalability-memory} plots these results.

% increase in launch latencies 
% is much less steep for \tool. In other words, \tool\ improves launch latencies 
% by \note{$\sim$$x\%$} over vanilla \docker.

% We also measure the daemon's memory utilization for both vanilla \docker\ and 
% \tool. Figure~\ref{figure:scalability-memory} plots the results. We observe 
% that for $500$ container launches, \tool's daemon utilizes \note{$\sim$$y\%$} 
% less memory than vanilla \docker. We attribute this difference to 
% \note{\todo{Fix it.}}. 

\subsection{Threats to validity}
\label{sec:evaluation:threats}

While the benchmarks used for evaluation are representative of the popular 
container applications used in practice, our evaluations were performed under a 
controlled setting within a virtual machine. It is possible that a combination 
of compute and/or memory intensive load might induce performance behaviors not 
captured in our study.

%% file: sections/relatedwork.tex
\section{Related Work}
\label{sec:relatedwork}

% \tool\ is most closely related to \code{rkt}~\cite{rkt} and LXD~\cite{lxd}. Both 
% these systems provide mechanisms to decouple the containers from the 
% orchestration platform. However, none of them is as well adopted as \docker. 

% \tool\ is architecturally closely related to \code{rkt}~\cite{rkt} and 
% runC~\cite{runc}. 
Like \tool, both \code{rkt}~\cite{rkt} and runC~\cite{runc}, ensure container 
reliability, and in the worst case, just one container is impacted. However, 
unlike \tool's light-weight monitor, both these tools attach a separate 
heavy-weight execution engine to each container, thereby  
increasing the system's memory footprint. Furthermore, \tool\ is compatible with 
the entire \docker\ eco-system, while \code{rkt} is still not widely adopted, 
and runC is under active development and lacks many useful functionalities.

LXD~\cite{lxd} is a container ``hypervisor'' to manage containers. It implements 
a REST API atop LXC~\cite{lxc}, a system container runtime, and forks a monitor 
and container process, which ensures that the LXD daemon is not a central point 
of failure and containers continue running in case of LXD daemon failure. Even 
though architecturally similar to \tool, LXD is closer in functionality to 
KVM~\cite{kvm}, a full system virtualization platform.

OpenVZ~\cite{openvz} is a system container runtime designed to execute full 
system images. In contrast, \code{systemd-nspawn}~\cite{nspawn} handles only 
process isolation and does not enable resource isolation. \tool\ on the other 
hand enhances reliability and availability of containers that leverage the 
\docker\ runtime.

% Lithos~\cite{lithos}, like \docker, is a base tool to build container 
% orchestration, but does not include image downloader, builder and other 
% networking capabilities. In contrast CoreOS~\cite{coreos} and Google's 
% Kubernetes~\cite{kubernetes} are fully functional orchestration platforms for 
% handling containers in cluster environments. However, none of them, unlike 
% \tool, address issues with container reliability.

%% file: sections/conclusion.tex
\section{Conclusion}
\label{sec:conclusion}

We present \tool, a system that re-architects the process hierarchy in \docker\ 
to significantly enhance reliability and availability of the containers, while 
imposing only moderate overheads. Container cloud offerings built atop \tool\ 
would enable a more reliable and available execution environment.